# Decision Theory with Degree of Membership in the State Space

Jeff Dominitz
Department of Economics and Ken Kennedy Institute
Rice University, Houston, TX 77005, USA
jeff.dominitz@rice.edu

Charles F. Manski
Department of Economics and Institute for Policy Research
Northwestern University, Evanston, IL 60208, USA
cfmanski@northwestern.edu

Abstract

Classical decision theory considers a decision maker (DM) who faces a choice set C and believes that the true state of nature $s^*$ lies in a state space S. An objective function $w(\cdot, \cdot): C \times S \to R^1$ maps actions and states into welfare. The DM wants to maximize $w(\cdot, s^*)$ but does not know $s^*$. The state space specifies a labeled set of possible states. Possibility is binary: every state in S is possible and no state outside of S is possible. Thinking that this binary perspective on possibility is too rigid, we generalize the classical theory. We begin with a universe of labeled states U and a residual unlabeled state r. Classical theory associates a binary possibility indicator $\mu(s)$ with each $s \in (U, r)$. The residual state r is not possible, so $\mu(r) = 0$. The state space is $S = [s \in U: \mu(s) = 1]$. We expand the range of $\mu$ to be the interval [0, 1]. Thus, $\mu$ is now a real number measuring *degree of membership*, a term we adopt from fuzzy set theory. The *membership function* $\mu(\cdot): (U, r) \to [0, 1]$ differs substantively from but is consistent with the subjective probability distributions of Bayesian decision theory. Bayesian theory places a subjective distribution on the elements of a classical state space, each of whose members has degree of membership one. Degree of membership expresses the distinct idea that a DM may want to quantitatively express uncertainty regarding the composition of the state space. A mathematical distinction between subjective probabilities and degrees of membership is that the former must sum to one, but the latter need not. After formalizing basic ideas, we extend three prominent classical decision criteria−maximization of subjective expected welfare, minimax regret, and maximin−to our generalization of classical decision theory.

## 1. Introduction

Savage (1954) used the term "small world" to describe the domain of classical decision theory. The theory considers a decision maker (DM) who faces a choice set C and who knows (or believes) that the true state of nature $s^*$ lies in a state space S. An objective function $w(\cdot, \cdot)$: $C \times S \Rightarrow R^1$ maps actions and states into welfare. The DM wants to maximize $w(\cdot, s^*)$ over C but does not know $s^*$.

The state space specifies a labeled set of possible states of nature, sometimes called a set of "known unknowns." Possibility is binary. Every state in S is possible and no state outside of S is possible. Decision theory does not explain how a DM determines possibility. The DM excludes the possibility that the truth may be some unlabeled residual state. Machina (2003) puts it this way (p. 30): "When the decision maker has reason to 'expect the unexpected,' . . . . the best one can do is specify a final catch-all state, with a label like 'none of the above,' and a very ill-defined consequence." This residual state may capture "unknown unknowns," outside the small world domain of classical decision theory.

Study of decision theory has had three distinct objectives: normative, descriptive, and prescriptive. Normative analysis seeks to establish ideal properties of decision making, aiming to give meaning to the concepts "optimal" and "rational." Descriptive analysis uses the theory to help understand and predict how actual decision makers behave. Prescriptive analysis seeks to improve the performance of actual decision making. Our concern is prescriptive. We want decision theory to be useful to decision making in practice, by enhancing the welfare achieved by a DM. This pragmatic objective differs from the normative orientation of axiomatic decision theory, which values certain forms of consistency of behavior across hypothetical choice settings as a virtue per se. See Manski (2011) for further discussion distinguishing prescriptive from normative decision theory.

Observing that abstract theory provides no guidance on how to specify the state space, we wrote the following in Dominitz and Manski (2026), a paper focused on decision making using artificial intelligence (p. 11):

"Broadly speaking, we might regard intelligence to be a general ability to specify a realistic state space. The state space should manifest neither incredible certitude, which assumes erroneous information about the environment, nor nihilism, which fails to use credible information. Decision making with a smaller than realistic state space erroneously classifies some actions as dominated. Use of a larger than realistic state space erroneously classifies some actions as undominated."

Unfortunately, this guidance is too loose. It does not suggest an operational way to proceed. Recognizing this, we wrote:

"The central open issue . . . . is interpretation of the words "credible" and "realistic." Simply stating that an intelligent DM should specify a credible or realistic state space is meaningless per se. It just transfers vagueness in the definition of intelligence to vagueness about the meaning of credibility and realism."

As we now view the matter, a practical problem in classical decision theory is that the binary perspective on possibility used in specifying a state space is too rigid. In this paper we extend the classical theory. The objective is to increase its applicability to the larger worlds that DMs actually inhabit.

We begin with a universe of labeled states U and a residual unlabeled state r. To simplify notation, we conceptualize r as a single state rather than a set of states; this does not affect our analysis. Within this framework, classical decision theory associates a binary possibility indicator $\mu(s)$ with each $s \in (U, r)$. The residual state r is not possible, so $\mu(r) = 0$. The state space is $S = [s \in U: \mu(s) = 1]$. The DM knows $[\mu(s), w(c, s); c \in C, s \in U]$. The DM does not know $[w(c, r), c \in C]$, but this is immaterial because $\mu(r) = 0$.

To go beyond classical decision theory, we expand the range of $\mu$ to be the interval [0, 1]. Thus, $\mu$ is now a real number measuring *degree of membership*. We adopt the term degree of membership from fuzzy set theory (Zadeh, 1965), which extends the classical concept of set membership from a binary logical statement to a real number in the interval [0, 1]. In Zadeh's language, the state space is a fuzzy set and $\mu(\cdot)$: $(U, r) \rightarrow [0, 1]$ is the *membership function*.

Specification of a state space S in classical decision theory is a subjective task performed by a DM. Specification of a degree of membership for each state in U is likewise a subjective task for a DM. It is natural to think that the membership function $\mu(\cdot)$ chosen by a DM depends on the information that the DM possesses. This information may include both knowledge obtained from theoretical studies and evidence from empirical research. For notational simplicity, we leave implicit in the notation $\mu(\cdot)$ the dependence of degree of membership on the DM's information.

*Note on Semantics*: We adopt Zadeh's term "degree of membership" because it has a neutral mathematical heritage that does not suggest a particular interpretation. We considered use of other terms that are more suggestive in everyday language, such as degree of "feasibility" or "possibility" or "credibility" or "realism." However, we concluded that each of these and related terms is inevitably loose and lacking in a consensus interpretation. For example, one of us has long struggled to meaningfully interpret the familiar word "credibility." Manski (2013) wrote (p. 12):

> "I will use the word *credibility* throughout this book, but I will have to take it as a primitive concept that defies deep definition. The second edition of the *Oxford English Dictionary* (*OED*) defines *credibility* as "the quality of being credible." The *OED* defines *credible* as "capable of being believed; believable." It defines *believable* as "able to be believed; credible." And so we come full circle." ■

One might be inclined to consider the vector $\mu(\cdot) \equiv [\mu(s), s \in (U, r)]$ of degrees of membership to be just a notational expansion of Bayesian subjective probabilities $\pi(\cdot) \equiv [\pi(s), s \in S]$ that each s with $\mu(s) = 1$ is the true state. However, we intend $\mu(\cdot)$ as a distinct concept that differs mathematically and substantively from $\pi(\cdot)$. We explain abstractly in Sections 1.1 and 1.2. We summarize the remainder of the paper in Section 1.3.

### 1.1. The Mathematical Difference between Degree of Membership and Subjective Probability

Let U be countable for expositional simplicity. The mathematical distinction between subjective probability and degree of membership is that $\pi(\cdot)$ must satisfy the adding-up constraint $\sum_{s \in (U, r)} \pi(s) = 1$, but $\mu(\cdot)$ need not. $\mu(\cdot)$ can take any value in the unit hypercube, provided that there exists some $s \in U$ such that $\mu(s) > 0$.[1]

This mathematical distinction between $\mu(\cdot)$ and $\pi(\cdot)$ manifests itself when one poses ceteris-paribus questions concerning decision making with alternative degrees of membership. Let s be a specified state. For alternative degrees of membership $\mu(s)$ and $\mu'(s)$, consider this question: Ceteris paribus, how would decision making change if the degree of membership $\mu(s)$ were instead to equal $\mu'(s)$?

This question is well-posed when considering degrees of membership, but not when considering subjective probabilities. Suppose one were to ask: Ceteris paribus, how would decision making change if the subjective probability $\pi(s)$ were instead to equal $\pi'(s)$? This question is not well-posed, because it is not possible to change $\pi(s)$ ceteris paribus. The adding-up condition for a probability measure implies that a change in $\pi(s)$ must be accompanied by commensurate changes in the probabilities of other states to enforce the constraint that the probabilities of all states sum to one.

When we take the range of $\mu(\cdot)$ to be the interval [0, 1], interpreting the lower bound $\mu(s) = 0$ as complete impossibility is fundamental to our analysis. In Sections 3 and 4, we maintain the classical assumption that $\mu(r) = 0$. This enables analysis that ignores the residual state. Section 5 confronts the problem of decision making when $\mu(r) > 0$.

Taking the upper bound on membership to be $\mu(s) = 1$ is not essential. The abstract aspects of our analysis developed in Section 3 only require that μ be ordinal. The more concrete aspects developed in Sections 4 and 5 place $\mu(\cdot)$ on a ratio scale with a finite upper bound. Making the upper bound equal one just simplifies notation.

[1] If U is a sufficiently large uncountable set, a further distinction arises. It may be formally possible only to define a probability measure that places all mass on a sufficiently "small" subset of U. This difficulty may seem esoteric, but it is a serious concern in nonparametric regression analysis. The issue does not arise with $\mu(\cdot)$, which need not be a measure.

## 1.2. The Substantive Difference between Degree of Membership and Subjective Probability

Proposing degree of membership as a mathematical construct would be of little interest in the absence of a substantive issue that the construct addresses. Our motivation in extending the concept of possibility from a binary indicator to degree of membership is to soften the rigid distinction in classical decision theory between states inside and outside the state space. In the classical theory, the state space is S = [s ∈ U: μ(s) = 1]. All other states in U have zero possibility, making them irrelevant to decision making. In the extended theory, all states s ∈ (U, r) with μ(s) > 0 may be relevant to decision making. States with complete membership μ(s) = 1 do not have a special status.

The subjective probabilities of Bayesian decision theory associate a quantitative *degree of likelihood* with each state in the state space. We have already observed that subjective probabilities differ from degrees of membership in having to add up to one. They also differ in their domain. For s ∈ S, π(s) is the likelihood that s is the true state of nature conditional on a maintained belief that the true state definitely lies in the prespecified state space S. Subjective probabilities are not defined for states outside of S. In contrast, μ(s) is defined for all s ∈ (U, r), where (U, r) exhaustively lists all possibilities.

Degree of membership should not be confused with extensions of Bayesian theory in which a DM contemplates more than one subjective distribution. These extensions, variously called *multiple prior* or *imprecise probability* or *robust Bayes* theory, suppose that the DM places a set of subjective distributions on the same pre-specified state space S ≡ [s ∈ U: μ(s) = 1] as does standard Bayesian analysis. They do not address a labeled universe U or a residual state r.[2]

[2] A DM who is committed to the idea that uncertainty should always be expressed through a subjective probability distribution may suggest interpretation of μ(s) as a DM's subjective probability that any s ∈ (U, r) is possible. To formalize this, the DM could retain the classical supposition that possibility is binary and view μ(·) as a subjectively random vector whose elements can take the values 0 and 1. The DM would thus place a subjective distribution φ(·) on all non-empty subsets V of (U, r), with φ(V) being the subjective probability that [μ(s) = 1, s ∈ V; μ(s) = 0, s ∈ (U, r)/V]. From this perspective, there is not a prespecified state space S but rather a subjective distribution of possible state spaces.

Although classical decision theory has generally not contemplated a residual state, an exception is economic research on incomplete contracts, which conceptualizes the residual state in various ways. In an article scrutinizing the foundations of the theory of incomplete contracts, Maskin and Tirole (1999) proposed placement of a subjective probability on a residual state composed of "unforeseen contingencies" and placement of a subjective distribution on payoffs in this state. In an article critiquing the proposal of Maskin and Tirole, Hart and Moore (1999) posed the issue as "indescribability" of states rather than as unforeseen contingencies. Although these two articles differed in multiple respects, they both used probabilistic concepts to study incompleteness using expected-utility theory.[3]

We do not pursue a probabilistic perspective on possibility because we think it essential for decision theory to encompass settings where a DM feels unable to form credible subjective probability distributions. To cope with these settings, classical decision theory has proposed and studied decision criteria that, in different senses, behaves uniformly well across the state space. Prominent such criterion are minimax-regret (MMR) and maximin (MM). Classical application of these criteria necessitates pre-specification of a state space. We extend MMR and MM to environments that lack a definitive state space.

### 1.3. Contributions of the Paper

The central new contribution of this paper is our proposal to extend classical decision theory to use the concept of degree of membership in the state space. Our objective is to enable greater realism in applications of decision theory, by softening specification of the state space. We are not aware of precedent research developing this idea. A further contribution of the paper is that we suggest a new way to use the idea of a residual state in decision making.

[3] An entirely different attempt to cope with the possibility of a residual state is made in econometric and statistical research on specification testing, as in Hausman (1978). Here the specified state space is a parametric model. Two parameter estimators are compared using frequentist hypothesis testing. Under the null hypothesis, both estimators are consistent and so converge to the same parameter value asymptotically. If the null hypothesis is not correct, the estimators have different probability limits asymptotically. In practice, if the test statistic rejects the null hypothesis, one tentatively concludes that the parametric model does not contain the true state of nature. Implications for decision making are not explored in this body of work.

Before commencing formal analysis in Section 3, Section 2 illustrates how applied research might use our new concepts. These initial illustrations draw on our past and ongoing applied research. They include policy choice using observational panel data, choice of climate policy under deep uncertainty, and choice of monetary policy under model uncertainty. Further illustrations that use the findings of our analysis are provided in Sections 4 and 5.

The analysis in Sections 3 and 4 maintains the conventional condition that no residual state exists. Section 3 suggests three properties that we think reasonable decision criteria using the concept of degree of membership should possess. Two properties are obvious; a decision criterion should respect weak dominance and should be unaffected by welfare in states that have zero degree of membership. The third property is more intriguing. We propose that if a decision criterion yields a choice that would be optimal if a specified state t were true, it should also yield that choice if state t were to have higher degree of membership, ceteris paribus.

Section 4 asks whether extensions of three prominent decision criteria−maximization of subjective expected welfare, minimax regret, and maximin−satisfy the reasonableness properties. In particular, we study criteria that weight state-specific welfare by the degree of membership of the state. We find that the weighted expected welfare and minimax regret criteria always satisfy all three properties, but weighted maximin often violates the second or third property. Given this, the remainder of the paper focuses on the weighted expected welfare and minimax regret criteria. We discuss how DMs applying these criteria might use sample data to inform beliefs, noting that a DM who maximizes weighted expected welfare may emulate Bayesian updating to obtain posterior membership weights that add up to one across states. By contrast, sample data should not affect degree of membership for a DM who applies weighted MMR within the ex ante framework of frequentist statistical theory. Section 4 closes with an illustration to clinical choice between surveillance and aggressive treatment when illness is uncertain.

Section 5 extends the analysis to permit a residual state with positive degree of membership. Decision theory has long struggled to cope with the possibility that a specified state space may not exhaust all possible true states of nature. We mentioned above the discussion in Machina (2003) and research on incomplete contracts. The literature review section in Karni and Viero (2017) references a large set of papers, which they categorize into three approaches.

It is obvious that any operational way to cope with a residual state must place some structure on the concept. Rather than place a subjective distribution on welfare in the residual state, we propose bounding welfare in the residual state, in which case the DM faces a problem of "partially known unknowns." We are not aware of previous work pursuing this idea. We show that the implications depend on the form of the bound and on the decision criterion used. We focus on bounded variation assumptions and use the *precautionary prin*ciple to illustrate.

Section 6 concludes by looking ahead towards empirical research to determine whether DMs find the concept of the degree of membership provides an appealing and useful generalization of classical decision theory.

## 2. Initial Illustrations

### 2.1 Policy Choice Using Observational Panel Data

Manski and Pepper (2013, 2018) considered the use of observational panel data to predict the mean treatment outcome that would occur if all individuals with specified observed covariates were to receive a counterfactual treatment at a specified date. This problem arises often in evaluation of public policy. The methodological analysis in these articles addressed controversial questions in criminal justice policy. The 2013 article concerned the effect on murder rates of laws permitting capital punishment and the 2018 one studied the effect on crime rates of laws permitting citizens to possess and carry handguns.

Abstracting from these applied settings, both articles supposed that each member of a population has one of two covariate values, x = v and x = w. At date d, each individual j receives one of two mutually exclusive and exhaustive treatments (A, B), which have potential outcomes, $y_{jd}(A)$ and $y_{jd}(B)$. The observed outcome for each individual j at each date d is $y_{jd} \equiv y_{jd}(z_{jd})$, where $z_{jd} = B$ if j receives B at d and $z_{jd} = A$ otherwise. Outcomes are observed at two dates, labeled d = 0 and d = 1. At d = 0, all members of the population receive A. At d = 1, those with x = v receive B and those with x = w receive A.

The problem is to learn $E[y_1(A)|x = v]$, the counterfactual mean outcome at d = 1 if individuals with covariate v were to receive A rather than B. Policy researchers often want to know this counterfactual mean in order to compare outcomes under policies that mandate treatments B and A respectively. At d = 1 these mean outcomes are $E[y_1(B)|x = v]$ and $E[y_1(A)|x = v]$ respectively. The available data reveal $E[y_1(B)|x = v] = E(y_1|x = v)$ but not $E[y_1(A)|x = v]$.

2.1.1. Conventional Analysis Using Invariance Assumptions

The conventional research practice when studying settings with this structure has been to invoke assumptions that are strong enough to point-identify the counterfactual mean outcome. These assumptions typically assert invariance of some kind, which enable use of the available data to yield three different findings on the counterfactual mean outcome, as follows:

(1) ***Time invariance***: $E[y_1(A)|x = v] = E[y_0(A)|x = v] = E(y_0|x = v)$;

(2) ***Cross-Covariate invariance***: $E[y_1(A)|x = v] = E[y_1(A)|x = w] = E(y_1|x = w)$;

(3) ***Difference-in-difference invariance***: $E[y_1(A)|x = v] =$

$$\{E[y_1(A)|x = w] - E[y_0(A)|x = w]\} + E[y_0(A)|x = v] = [E(y_1|x = w) - E(y_0|x = w)] + E(y_0|x = v).$$

Assumption (1) is correct under the assumption that no determinant of outcomes changes for individuals with covariate v between d = 0 and d = 1 except for replacement of treatment A by B. Assumption (2) is correct under the assumption that individuals with covariate v and w respond to treatment in the same way. Assumption (3) is correct under the assumption that, if individuals with covariate v had received treatment A at d = 1, they would have experienced the same mean change in outcomes between d = 0 and d = 1 as those with covariate w experienced.

Thus, each of three findings is obtained with a specific invariance assumption. It is generally the case in practice that these invariance assumptions do not hold jointly. Each assumption is strong and may be unrealistic. It may be that none of them holds.

### 2.1.2. Bounded Variation Assumptions

The above invariance assumptions have a sharpness that often makes them suspect. Each of the three equalities (1), (2), and (3) point-identify the counterfactual mean outcome $E[y_1(A)|x = v]$. Why should any of them hold exactly? Empirical researchers often say that such assumptions are "approximations," but they do not formalize what this means.

Manski and Pepper observed that a simple way to improve the credibility of empirical research is to weaken invariance assumptions to assumptions of *bounded variation*. They weakened assumptions (1), (2), and (3) as follows:

(4) ***Bounded Time Variation***: $|E[y_1(A)|x = v] - E[y_0(A)|x = v]| \leq \delta_{v(10)}$;

(5) ***Bounded Cross-Covariate Variation***: $|E[y_1(A)|x = v] - E[y_1(A)|x = w]| \leq \delta_{(vw)1}$;

(6) ***Bounded Difference-in-difference Variation***:

$$|\{E[y_1(A)|x = w] - E[y_0(A)|x = w]\} - \{E[y_1(A)|x = v] - E[y_0(A)|x = v]\}| \leq \delta_{(vw)(10)};$$

The δ values are specified positive constants.

Each bounded variation assumption has identifying power, whose strength is determined by the magnitude of the relevant δ constant. A DM can in principle set one, two, or all three constants to infinity.

Hence, we can without loss of generality view (4)-(6) as a collection of assumptions that may be imposed in combination.

When all three constants are positive and at least one is finite, the combination of assumptions may partially identify the counterfactual mean outcome, yielding an informative interval that contains $E[y_1(A)|x = v]$. If more than one constant is positive and at least one is close to zero, it may be that no value of $E[y_1(A)|x = v]$ satisfies the combination of assumptions. If so, the DM should conclude that some part of the asserted combination must be incorrect.

2.1.3. Analysis with Degree of Membership in the State Space

The state space in conventional analysis using invariance assumptions (1), (2), or (3) is a singleton, the specific point depending on which assumption is made. A minor generalization would be for a DM to assume that one of the three assumptions is correct, without knowing which one. Then the state space contains three points.

Assumptions (4), (5), and (6) express a broad spectrum of generalizations that bound $E[y_1(A)|x = v]$. The width and location of the bound depends on the values assumed for the three $\delta$ constants, thereby expanding the state space to a continuum of possibilities. Nevertheless, analysis remains within the scope of classical decision theory. Whatever values are assumed for the $\delta$ constants, some values of $E[y_1(A)|x = v]$ are assumed possible and others are not.

Analysis with degree of membership would soften this sharp distinction. A DM might suppose that a set of $E[y_1(A)|x = v]$ values formed with small $\delta$ constants has a low degree of membership. One formed with a larger $\delta$ has a higher degree of membership.

The idea of a residual state is unnecessary in this illustration. The value of $E[y_1(A)|x = v]$ must lie on the real line. Hence, there is no need to conjecture unlabeled states.

## 2.2. Choice of Climate Policy under Deep Uncertainty

Climate economists use *integrated assessment* (IA) models to evaluate the benefits and costs of alternative climate policies. Policy comparisons are performed by considering a planner who seeks to make optimal trade-offs between the costs of greenhouse gas (GHG) abatement and the damages from climate change.

Economists have formalized policy choice as an optimal control problem with these components: (1) equations coupling GHG emissions and abatement to the accumulation of GHGs in the atmosphere and resulting temperature increases; (2) a damage function that quantifies effects of climate change in terms of the loss of global economic output as a function of temperature increases; (3) an abatement cost function that expresses the cost of actions to reduce GHG emissions relative to a stipulated baseline emissions trajectory. A prominent example is the DICE (Dynamic Integrated Climate Economy) model (Nordhaus, 2019).

Studying climate policy as a control problem presumes that a planner knows enough to make optimization feasible, but uncertainties abound. Physical scientists have performed multi-model ensemble (MME) analysis. Lacking a consensus climate model, they have developed multiple models, each assuming that a different model accurately describes the climate system. They average the forecasts of the models in an MME. Taking a simple average is often called "model democracy." Weighted averages may be called Bayesian model averaging; see Smith *et al*. (2009).

Economists have estimated various damage functions and abatement cost functions. In general, economists have not performed MME analyses that combine multiple functions by averaging. Instead, each study specifies a parametric family of functions and assumes the specification to be correct. Thus, each researcher assumes that the state space is a parametric family.

### 2.2.1. Minimax-Regret Analysis of Climate Policy

Concerned with the limited credibility of research viewing climate policy choice as an optimal control problem, Manski, Sanstad, and DeCanio (2021) and DeCanio, Manski, and Sanstad (2002) framed uncertainty in climate modeling as a problem of partial identification, generating *deep uncertainty*. It was observed that climate physics, the damages from global warming, and abatement costs are incompletely known and there is no basis to choose one among the multiple models that have been proposed. These articles proposed use of the minimax-regret decision criterion to account for this deep uncertainty. They developed a theoretical framework for cost-benefit analysis of climate policy based on MMR and applied it computationally with a simple illustrative IA model, which supposed that the correct climate model is one of six prominent models in the literature on climate science.

### 2.2.2. Analysis with Degree of Membership in the State Space and a Residual State

Conventional IA analysis has assumed unrealistically precise knowledge of climate physics and economics. The MMR analyses described above assume a somewhat larger state space but do not claim to be fully realistic. There is no justification to think that one of the six climate models in the assumed state space is correct. Many other climate models have been studied, each yielding its own forecasts. It may be that none of the models proposed to date is fully accurate.

Analysis with degree of membership and a residual state would enable a planner to express a more nuanced perspective on current knowledge. The planner could assign a degree of membership to each model proposed to date, without assuming that one of these models must be correct. The possibility that no existing model is correct could be permitted by allowing the possibility that the truth lies in a thus far unspecified residual state.

## 2.3. Choice of Monetary Policy under Model Uncertainty

Economists have also formalized monetary policy choice as an optimal control problem. In the U.S., the Federal Reserve faces the so-called dual mandate to maintain maximum employment and stable prices. Macroeconomic research has long framed the dual mandate in terms of quadratic loss in deviations of aggregate economic activity (either GDP or unemployment) and inflation from target levels (e.g., Barro and Gordon, 1983; Rogoff, 1985). Quadratic loss minimization has been interpreted as an approximation to social welfare maximization near the steady state (Woodford, 2010). The central bank chooses the path of short-term interest rates to minimize expected loss subject to a macro model of economic dynamics.

Numerous models of the economy have been proposed and used to evaluate policy, often taking the form of a dynamic stochastic general equilibrium (DSGE) model. Since 1996, the Federal Reserve has used and updated the FRB/US model, described as a variant of a DSGE model that "applies optimization theory more flexibly."[4] Each such model, DSGE or otherwise, represents a different point in the state space. The FRB/US model allows for two states of nature, one in which agents form model-consistent expectations of future outcomes and one in which agents use vector autoregression to form expectations.

Some research has considered how a central bank does or should account for model uncertainty in conducting monetary policy, including Levin and Williams (2003), Brock et al. (2007a), and Brock et al. (2007b). In seeking to describe why the Federal Reserve was slow to implement a low-inflation policy in the 1970s, Cogley and Sargent (2005) modeled the central bank as a Bayesian decision maker performing model averaging over three possible states labelled as a Samuelson-Solow model, a Solow-Tobin model, and a Lucas-Sargent model. No residual state was considered. However, with a period of analysis beginning in 1960, priors on each state had to be assigned as of Q4 in 1959, predating the research developing the Solow-Tobin and Lucas-Sargent models. Cogley and Sargent acknowledged this concern and assigned priors of 0.01 to each of the two models. These models would more appropriately have been described as elements of a residual state space at that time.

[4] See https://www.federalreserve.gov/econres/us-models-about.htm, accessed October 3, 2026.

Hansen and Sargent (2001a, 2001b) took a qualitatively different approach to monetary policy with model uncertainty. Rather than applying optimal control theory with a single model assumed to be correct, they applied robust control theory with a single approximating model and a bounded variation assumption on the state space. In particular, they expanded the state space by allowing for what Cogley and Sargent (2005) subsequently called (p. 533) "an uncountable cloud of alternative models" with entropy that is bounded relative to the approximating model. Further, Hansen and Sargent applied a minimax decision rule, thereby sparing the decision maker of the need to specify priors over that uncountable cloud.

## 3. Reasonable Decision Criteria

We consider decision criteria that map [μ(s), w(c, s); c ∈ C, s ∈ (U, r)] into a real-valued function f(·). Given specified values for degrees of membership and welfare, this function expresses a preference ordering on C. The DM chooses an action that maximizes or minimizes f over C, the relevant extremum depending on the function. The present discussion supposes that preference increases in f(·). The discussion may be rephrased for preference decreasing in f(·).

The one essentially universally accepted principle of reasonableness in classical decision theory is that a decision criterion should respect weak dominance. Extending this idea to our extended framework, we say that f(·) respects weak dominance if it satisfies this requirement:

*Respect for Weak Dominance (RWD)*: Let μ(r) = 0. Consider actions (d, e) such that w(d, s) ≥ w(e, s) for all s ∈ U with μ(s) > 0 and w(d, s) > w(e, s) for some such s. Then, f(d) ≥ f(e). ■

This definition does not cover settings with $\mu(r) > 0$. In the absence of any knowledge of $[w(c, r), c \in C]$, no action pairs (d, e) are known to satisfy the premise $w(d, r) \geq w(e, r)$. Section 5 will consider settings with $\mu(r) > 0$ and partial knowledge of welfare in state r.

Characterization of reasonableness in the extended framework necessitates consideration of how a decision criterion should behave as a function of $\mu$. Heuristically, we should want states with higher degrees of membership to influence choice more than states with lower degrees of membership. The question is how to formalize this idea.

We think a basic reasonableness property to be that choice should not be affected by welfare in states with zero membership. The state space S of classical decision theory is $[s \in U: \mu(s) = 1]$. Extending this, we henceforth let $S \equiv [s \in U: \mu(s) > 0]$. We let U/S denote the complementary subset of U with zero membership. We propose this reasonableness property:

*Irrelevance of States with Zero Membership (ISZM)*: Ceteris paribus, $f(\cdot)$ does not vary with $[w(\cdot, s), s \in U/S]$. ■

A further property that we think prima facie reasonable emerges by fixing $[w(\cdot, s), s \in S]$ and comparing the choices that a DM makes as the membership vector $[\mu(s), s \in S]$ varies. We formalize this broad idea in a particular way, using as a benchmark the choice that the DM would make if some state $t \in S$ were known to be true. In this hypothetical setting, a rational DM would choose an element of $C[W(t)] = \text{argmax}_{c \in C}\, w(c, t)$, where $W(t) \equiv [w(c, t); c \in C]$. We use the set notation $C[W(t)]$ rather than the point notation $c[W(t)]$ to express the possibility that $W(t)$ may have multiple maximal elements.

Consider decision criterion $f(\cdot)$ and membership vector $[\mu(s), s \in S]$. Denote the chosen action using $f(\cdot)$ as $c[\mu(t), W(t), V(t)] \in \text{argmax}_{c \in C}\, f[c; \mu(t), W(t), V(t)]$, where $V(t) \equiv [\mu(s), w(c, s); c \in C, s \in S, s \neq t]$. We write that the chosen action is an element of the argmax because f may have multiple maximizers. Now hold $[W(t), V(t)]$ fixed and consider how $c[\mu(t), W(t), V(t)]$ varies with the scalar $\mu(t)$. We propose this reasonableness property.

*Respect for Degree of Membership (RDM)*: Consider any $[t, \mu(t), W(t), V(t)]$ with $\mu(t) < 1$. Let $\mu^+(t) > \mu(t)$.

A. If $c[\mu(t), W(t), V(t)] \in C[W(t)]$, then $c[\mu^+(t), W(t), V(t)] \in C[W(t)]$.

B. If $c[\mu^+(t), W(t), V(t)] \notin C[W(t)]$, then $c[\mu(t), W(t), V(t)] \notin C[W(t)]$. ■

Parts A and B are logically equivalent, as Part B is the contrapositive restatement of A. In words, part A states that if criterion $f(\cdot)$ yields a choice that would be optimal if the true state were known to be t, it should also yield such a choice if t were to have higher membership, ceteris paribus.

We do not propose properties beyond RWD, ISZM, and RDM as having compelling claims to reasonableness. We leave open whether others will emerge.

## 4. Extending Prominent Classical Decision Criteria

Three prominent classical decision criteria are maximization of subjective expected welfare, minimax regret, and maximin. In the first, the DM places a subjective probability distribution $\pi$ on the classical state space S, averages state-dependent welfare with respect to $\pi$, and maximizes subjective expected welfare. The criterion solves

(7) $$\max_{c \in C} \sum_{s \in S} \pi(s)w(c, s).$$

The MMR and MM criteria, which do not assert a subjective probability distribution, are

(8) $$\min_{c \in C} \max_{s \in S} [\max_{d \in C} w(d, s) - w(c, s)].$$

$$\text{(9)} \quad \max_{c \in C} \; \min_{s \in S} \; w(c, s).$$

We extend these criteria in a simple manner that we think worthy of investigation. In each case, we replace welfare by the product of welfare and degree of membership. We extend the classical criteria in this way to formalize the broad idea that states with higher degrees of membership should influence choice more than states with lower degrees of membership. We ask whether the reasonableness properties of Section 3 hold. We find that they do for the extensions of criteria (7) and (8). However, the extension of criterion (9) violates the ISZM and RDM properties.

In this section, we maintain the condition $\mu(r) = 0$, leaving examination of settings with $\mu(r) > 0$ to Section 5. We formally consider all states in the universe U rather than the subset S of states with positive membership, but this is immaterial to the analysis.

4.1. Maximization of the Cross-State Sum of Membership-Weighted Welfare

Disregarding its interpretations in cognitive or axiomatic theories of subjective expected utility maximization, the algebraic structure of (7) is maximization of a weighted sum of welfare across states, all weights being non-negative and some being positive. We replace subjective probability weights with membership weights, giving the criterion

$$\text{(10)} \quad \max_{c \in C} \; \sum_{s \in U} \mu(s)w(c, s).$$

The algebraic distinction between (10) and (7) is immaterial if $\sum_{s \in U} \mu(s)$ is finite. When this holds, division of (10) by $\sum_{s \in U} \mu(s)$ and use of the fact that $\mu(s) = 0$ for $s \in U/S$ yields (7) with $\pi(s) \equiv \mu(s)/[\sum_{s \in S} \mu(s)]$. The distinction is material if U has infinite cardinality and $\sum_{s \in U} \mu(s)$ is infinite. Then criterion (10) is well-posed for some vectors $[\mu(s), w(c, s); c \in C, s \in U]$ but not for others. When (10) is well-posed, a Bayesian might be inclined to interpret it as maximization of expected welfare using an *improper prior*.

Inspection of criterion (10) shows that it satisfies properties RWD and ISZM. We show below that the criterion satisfies property RDM.

Proof of Property RDM: The premise is that, for a specified state t, action $c[\mu(t), W(t), V(t)]$ maximizes $w(\cdot, t)$ on C. Ceteris paribus, increasing the membership of state t from $\mu(t)$ to $\mu^+(t)$ adds $[\mu^+(t) - \mu(t)]w(c, t)$ to the sum in (10) for each $c \in C$, where $\mu^+(t) - \mu(t) > 0$. Hence, $[\mu^+(t) - \mu(t)]w(\cdot, t)$ is maximized at $C[W(t)]$. It follows that the action $c[\mu^+(t), W(t), V(t)]$ chosen with membership $\mu^+(t)$ is an element of $C[W(t)]$. ■

4.1.1. Bayesian Updating of Membership Weights with Sample Data

Consider a DM who observes sample data as part of the information he possesses. In classical decision theory, the subjective probability distribution used to maximize subjective expected welfare is obtained by specification of a distribution $\pi(\cdot)$ prior to observation of the sample data and application of Bayes Theorem to update the prior into a posterior distribution. Let $\psi$ denote the sample data and $L(\psi|s)$ be the likelihood of $\psi$ in state s. The posterior probability of s is $\pi(s|\psi) = L(\psi|s)\pi(s)/[\sum_{t \in S} L(\psi|t)\pi(t)]$.

Suppose that a DM who maximizes membership-weighted welfare observes sample data. How might this DM behave? A natural possibility is that the DM emulates Bayesian updating to obtain posterior membership weights for states of nature. Replacing the prior probabilities $\pi(\cdot)$ by prior membership weights $\mu(\cdot)$, Bayes Theorem yields posterior membership weights $\mu(s|\psi) = L(\psi|s)\mu(s)/[\sum_{t \in S} L(\psi|t)\mu(t)]$. Provided that the sum $\sum_{t \in S} L(\psi|t)\mu(t)]$ is finite, the posterior membership weights add up to one across states. Thus, a DM who a priori places membership weights on states behaves a posteriori as would a Bayesian DM.

4.2. Membership-Weighted Minimax Regret

The membership-weighted minimax-regret criterion solves the problem

$$(11)\quad \min_{c \in C}\ \max_{s \in U}\ [\max_{d \in C} \mu(s)w(d, s) - \mu(s)w(c, s)].$$

Equivalently, the criterion is

$$(11')\quad \min_{c \in C}\ \max_{s \in U}\ \mu(s)\,[\max_{d \in C} w(d, s) - w(c, s)].$$

The latter expression interprets the criterion as weighting regret rather than welfare. The larger the degree of membership of a state, the more this state is capable of being the one at which maximum weighted regret occurs.

Criterion (11)-(11′) reduces to classical MMR when μ(s) is constant for all states with positive membership, but it differs otherwise. Thus, membership-weighted MMR generically differs from classical MMR. This contrasts from maximization of the cross-state sum of membership-weighted welfare, which is generically equivalent to classical expected utility maximization.

Inspection of (11)-(11′) shows that it satisfies properties RWD and ISZM. We show below that it satisfies property RDM.

Proof of Property RDM: The premise of the property is that, for a specified state t, action c[μ(t), W(t), V(t)] maximizes w(·, t) on C. Hence, regret at t is zero when evaluated at this action. Hence, membership-weighted regret at t is zero when regret is multiplied by any membership value. A ceteris paribus increase in membership from μ(t) to $\mu^+(t)$ weakly increases regret in state t for actions other than c[μ(t), W(t), V(t)] and does not affect regret in other states. It follows that the action c[$\mu^+(t)$, W(t), V(t)] chosen with membership $\mu^+(t)$ is an element of C[W(t)]. ■

4.2.1. Membership-Weighted Minimax Regret with Sample Data

Consider again a DM who observes sample data. Whereas Bayesian decision theory contemplates choice of an action after observation of the data, Wald's statistical decision theory considers ex ante choice of a procedure in which the decision will depend on the realization of as yet unobserved sample data. That is, the DM must commit to a decision rule before seeing the data.[5]

The Wald theory lies within the general framework of frequentist statistical theory. The central concept of frequentist theory is the distribution of data across repeated samples. This is an ex ante concept. Evaluation of statistical procedures after observing finite-sample data does not occur in frequentist theory. Whether the focus of study be hypothesis testing, point estimation, construction of confidence sets, or decision making, frequentist theory adopts an ex ante perspective.

Classical analysis of MMR decision making with sample data applies the Wald theory to choose a decision procedure in a particular way. It replaces criterion (8) with

$$(8') \qquad \min_{c(\cdot) \in \Gamma} \ \max_{s \in S} \ \left( \max_{d \in C} w(d, s) - E_s\{w[c(\psi), s]\}\right).$$

Here $c(\cdot)$: $\Psi \rightarrow C$ is a statistical decision function that maps the sample data into a chosen action and $\Psi$ is the sample space of potential data realizations. Wald proposed measurement of the performance of $c(\cdot)$ in state s by its frequentist expected welfare across samples; that is, $E_s\{w[c(\psi), s]\} \equiv \int w[c(\psi), s]dQ_s$, where $Q_s$ is the sampling distribution in state s.

A natural extension of membership-weighted MMR to settings with sample data replaces (11′) by

[5] This ex ante perspective is appropriate in applications where a decision procedure must actually be chosen with commitment before observing data. There are many such cases. These include the design of algorithms that will be used contingent on the realization of future data. In algorithm design, the actual decision problem is to choose a decision rule before one observes sample data.

Whether the ex ante perspective is reasonable when a DM does not actually need to commit ex ante has been controversial. Then it seems heuristically sensible to wait to see the data before making a decision. Bayesian theory studies decision making without commitment, using Bayes Theorem to transform a prior subjective probability distribution into a posterior one after observing data. Bayesians argue that, when commitment is not required, it is not reasonable to evaluate procedures based on what decisions they would yield with sample data that ex ante could occur but ex post do not occur.

$$(11'') \qquad \min_{c(\cdot) \in \Gamma} \; \max_{s \in U} \; \mu(s) \left( \max_{d \in C} w(d, s) - E_s\{w[c(\psi), s]\} \right).$$

Consistent with the ex ante perspective of the Wald theory, this criterion does not use observed sample data to update the degree of membership function. It contemplates choice of a decision procedure before data are observed.

4.3. Membership-Weighted Maximin

The membership-weighted maximin criterion solves the problem

$$(12) \qquad \max_{c \in C} \; \min_{s \in U} \; \mu(s)w(c, s).$$

This extension of the classical maximin criterion (9) appears analogous to the extensions of (7) and (8) to (10) and (11-11′), but it has different properties. Criterion (12) satisfies property RWD, but it sometimes does not satisfy property ISZM or RDM.

The extended MM criterion does not satisfy property ISZM if $w(c, s) > 0$ for all $c \in C$ and $s \in U$. Consider any state $s_0$ with $\mu(s_0) = 0$ and any another state $s_+$ with $\mu(s_+) > 0$. Then, for each action c, $\mu(s_0)w(c, s_0) = 0$ and $\mu(s_+)w(c, s_+) > 0$. Hence, minimum membership-weighted welfare across U equals zero. This shows that states with zero membership are relevant to choice. Indeed, all actions solve the extended MM problem if such states exist.

In contrast to the above, the extended MM criterion satisfies property ISZM if $w(c, s) < 0$ for all $c \in C$ and some state with positive degree of membership. Again consider any state $s_0$ with $\mu(s_0) = 0$ and any another state $s_+$ with $\mu(s_+) > 0$. Then, for each action c, $\mu(s_0)w(c, s_0) = 0$ and $\mu(s_+)w(c, s_+) < 0$. Hence, minimum membership-weighted welfare across U is negative and occurs in a state with positive degree of membership. This shows that states with zero membership are irrelevant to choice.

An example suffices to show that the extended MM criterion can violate property RDM. In this example, we ensure property ISZM by supposing that w(c, s) < 0 for all c ∈ C and s ∈ U. Recall that Part A of the property states: If c[μ(t), W(t), V(t)] ∈ C[W(t)], then c[$\mu^+$(t), W(t), V(t)] ∈ C[W(t)].

*Example*: Suppose that C and U are binary, being C = {c, c′} and U = {s, s′}. Suppose further that:

$$\rho(s)w(c',s) < \rho(s)w(c,s) < \rho(s')w(c,s') < \rho(s')w(c',s').$$

Then c is the unique membership-weighted maximin choice and c is the unique optimal action if the true state were known to be s. Hence, the premise of condition A is satisfied. However, increasing $\rho(s)$ to $\rho^+(s)$ could yield

$$\rho(s')w(c,s') < \rho^+(s)w(c',s) < \rho^+(s)w(c,s) < \rho(s')w(c',s').$$

Then the worst state for c is s′, yielding membership-weighted welfare $\rho(s')w(c,s')$. The worst state for c′ is s, yielding weighted welfare $\rho^+(s)w(c',s)$. The latter is larger than the former, so c′ is now the membership-weighted maximin choice, violating the conclusion of Part A. ■

4.4. Illustration: Clinical Choice Between Surveillance and Aggressive Treatment when Illness is Uncertain

To add concreteness to the above abstract analysis, we now consider the prominent clinical decision problem of choice between surveillance and aggressive treatment when illness is uncertain. Early study of this problem determined optimal treatment assuming a singleton state space where the clinician knows the objective probability of illness for a patient with specified observed attributes (Pauker and Kassirer,1975, 1980; Phelps and Mushlin, 1988). Recent research has studied minimax-regret treatment assuming an interval state space where the clinician bounds the objective probability of illness (Manski, 2018).

Here we extend the latter analysis to permit possible values of the illness probability to vary in their degree of membership in the state space. We consider membership-weighted maximization of subjective expected welfare and minimax regret. We do not consider membership-weighted maximin in light of the unpalatable properties found in Section 4.3. In what follows, Section 4.4.1 summarizes the simple analysis of optimal decision making when the illness probability is known. Section 4.4.2 examines decision making when this assumption is weakened. Section 4.4.3 discusses specification of degrees of membership in practice.

4.4.1. Optimal Treatment Choice

Let there be two care options, denoted A and B, for a particular disease. Option A is surveillance and B is aggressive treatment. A is noninvasive and B is invasive. A clinician must choose one treatment or the other.[6]

The clinician chooses a treatment without knowing a patient's illness status; $y = 1$ if a patient is ill and $y = 0$ if not. The patient has observed health and demographic attributes, which we keep implicit for notational simplicity. Ideally, the clinician can learn the objective probability of illness $p \equiv P(y = 1)$.

Let expected patient welfare with option $c \in \{A, B\}$ have the known form $U(c, y)$; thus, the expected welfare of each option may vary with whether the disease occurs. It is usually reasonable to suppose that aggressive treatment is better if the disease occurs and surveillance is better otherwise. That is,

$$(13) \quad U(B,1) > U(A,1), \quad U(A,0) > U(B,0).$$

We assume that the chosen care option does not affect whether the disease occurs; hence, a patient's illness probability is simply p rather than a function p(c) of the care option. Treatment choice still matters because it affects the severity of illness and the patient's experience of side effects. Aggressive treatment is beneficial to the extent that it lessens the severity of illness, but harmful if it yields side effects that do not occur with surveillance.

---

[6] Assuming choice between these options disallows the possibility of randomized treatment choice of a single patient. It also disallows diversification when a health planner treats a population of observationally identical patients and can assign a fraction of patients to each treatment. Manski (2009) studies these settings.

Let the clinician know [p, U(·, ·)] and maximize expected welfare. Then an optimal decision is

(14) Choose A if $p \cdot U(A, 1) + (1 - p) \cdot U(A, 0) \geq p \cdot U(B, 1) + (1 - p) \cdot U(B, 0)$,

Choose B if $p \cdot U(B, 1) + (1 - p) \cdot U(B, 0) \geq p \cdot U(A, 1) + (1 - p) \cdot U(A, 0)$.

The decision yields optimal expected patient welfare

(15) $\max [p \cdot U(A, 1) + (1-p) \cdot U(A, 0), p \cdot U(B, 1) + (1 - p) \cdot U(B, 0)]$.

The optimal decision is easy to characterize when inequalities (13) hold. Let $p^{\#}$ denote the threshold value of p that makes options A and B have the same expected welfare. This value is

(16) $$p^{\#} = \frac{U(A, 0) - U(B, 0)}{[U(A, 0) - U(B, 0)] + [U(B, 1) - U(A, 1)]}$$

Option A is optimal if $p \leq p^{\#}$ and B if $p \geq p^{\#}$.

An instructive simplifying case occurs when aggressive treatment neutralizes disease, in the sense that $U(B, 0) = U(B, 1)$.[7] Let $U_B$ denote welfare with aggressive treatment. Then (13) – (16) reduce to

(17) $U(A, 0) > U_B > U(A,1)$.

(18) Choose A if $p \cdot U(A, 1) + (1 - p) \cdot U(A, 0) \geq U_B$,

Choose B if $U_B \geq p \cdot U(A, 1) + (1 - p) \cdot U(A, 0)$.

[7] For example, aggressive treatment might be surgery to remove a localized tumor that may (y = 1) or may not (y = 0) be malignant. Suppose that surgery always eliminates cancer when present. Then surgery neutralizes disease. Being invasive and costly, performance of surgery has a negative side effect on welfare that is the same regardless of whether cancer is present.

(19) $\max [p \cdot U(A, 1) + (1 - p) \cdot U(A, 0), U_B]$.

(20) $$p^{\#} = \frac{U(A,0) - U_B}{U(A, 0) - U(A, 1)}.$$

Further simplification occurs if one normalizes the location and scale of welfare by setting $U(A, 0) = 1$ and $U(A, 1) = 0$.[8] Then $1 > U_B > 0$. The clinician chooses A if $1 - p \geq U_B$ and B if $U_B \geq 1 - p$. Optimal welfare is $\max (1 - p, U_B)$. The threshold illness probability is $p^{\#} = 1 - U_B$. We henceforth assume that aggressive treatment neutralizes disease and normalize the welfare of surveillance in this manner.

4.4.2. Treatment Choice without Knowledge of the Illness Probability

A clinician who does not know whether p is smaller or larger than the threshold $p^{\#} = 1 - U_B$ faces a problem of decision making under ambiguity. Maximization of subjective expected welfare (7) is simple because expected welfare is linear in p. Hence, the clinician acts as if the subjective mean probability $\int p_s d\pi(s)$ equals the true value of p. Maximization of the cross-state sum of membership-weighted welfare is equally simple if $\sum_{s \in U} \mu(s)$ is finite. Then division of (10) by $\sum_{s \in U} \mu(s)$ and use of the fact that $\mu(s) = 0$ for $s \in U/S$ yields (7) with $\pi(s) \equiv \mu(s)/[\sum_{s \in S} \mu(s)]$. Hence, the treatment choice that maximizes membership-weighted welfare is B if

(21) $\sum_{s \in S} \mu(s)(1 - p_s) \leq U_B$

Option A maximizes membership-weighted welfare if the opposing inequality holds.

[8] This normalization is used in health economics research on *quality of life*, where the welfare value 1 is interpreted as perfect health and the value 0 as health so poor that a person would be indifferent between being alive and dead.

Manski (2018) studied classical MMR decision making when the clinician knows that p lies in an interval $p_m < 1-U_B < p_M$, where $p_m \equiv \min_{s \in S} p_s$ and $p_M \equiv \max_{s \in S} p_s$. The derivation is simple. Choosing A is an error in state s if $p_s > 1 - U_B$, yielding regret $p_s - (1 - U_B)$. MR is $p_M - (1-U_B)$, occurring when p attains its maximum possible value. Choosing B is an error if $p_s < 1 - U_B$, yielding regret $(1 - U_B) - p_s$. Maximum regret is $(1 - U_B) - p_m$, occurring when p attains its minimum value. Hence, the treatment choice that minimizes maximum regret is B if $(1 - U_B) - p_m < p_M - (1 - U_B)$, equivalently if $(p_m + p_M)/2 > 1 - U_B$. Symmetrically, A minimizes maximum regret if $(p_m + p_M)/2 < 1 - U_B$. MMR equals $\min\{p_M - (1 - U_B), (1 - U_B) - p_m\}$.

Minimization of membership-weighted MMR extends this derivation. Let $S_B$ be the subset of S where $p_s > 1 - U_B$. Choosing treatment A in any such state is an error, with membership-weighted regret $\mu(s)[p_s - (1 - U_B)]$. Maximum regret in $S_B$ is $\max_{s \in S(B)} \mu(s)[p_s - (1-U_B)]$. Symmetrically, let $S_A$ be the subset of S where $p_s < 1 - U_B$. Choosing B in any such state is an error, with weighted regret $\mu(s)[(1 - U_B) - p_s]$. Maximum regret in $S_A$ is $\max_{s \in S(A)} \mu(s)[(1 - U_B) - p_s]$. Hence, the treatment choice that minimizes maximum regret is B if

(22) $$\max_{s \in S(A)} \mu(s)[(1 - U_B) - p_s] \leq \max_{s \in S(B)} \mu(s)[p_s - (1 - U_B)].$$

Option A minimizes maximum regret if the opposing inequality holds.

The treatment choice that minimizes weighted maximum regret may or may not be the same as the one minimizing classical maximum regret. The choices obviously coincide if degree of membership is constant across states. Otherwise, the weighted MMR choice depends on the joint variation of $[\mu(s), p_s]$ across states.

4.4.3. Specifying Degrees of Membership in Clinical Practice

Specification of degrees of membership in practice requires clinical judgment. This necessarily depends on the medical context under consideration. The literature on Bayesian treatment choice supposes that clinicians place prior subjective distributions on illness probabilities and update them following realization of sample data in randomized trials and observational data collection. We expect that advocates of Bayesian treatment choice should be willing to contemplate placing prior degrees of membership on illness probabilities and using Bayes Theorem to update them, as discussed in Section 4.1.1.

Statistical analysis in medicine has mainly used frequentist rather than Bayesian concepts. The prevalent practice in computation of confidence intervals for illness probabilities has taken the state space for p to be the entire [0, 1] interval rather than some subset thereof. We do not have a ready explanation for this practice. Our informal observation of clinicians suggests that they rarely believe all values of illness probabilities to be equally possible. Perhaps practicing clinicians would want to specify a positive degree of membership to all illness probabilities, but we conjecture that they would rarely specify the same degree of membership to all $p \in [0, 1]$ and across patients with different attributes.

## 5. Analysis When the Residual State has Positive Degree of Membership

We now expand the scope of analysis to settings where the residual state has positive degree of membership. We have found that the weighted maximin criterion (12) sometimes does not satisfy the reasonableness properties of Section 3. Hence, we focus on the weighted extensions (10) and (11)-(11′) of maximization of subjective expected welfare and minimax regret.

Classical decision theory expresses all uncertainty as lack of knowledge of the true state of nature. For each possible state s, the welfare vector $[w(c, s), c \in C]$ is known. We cannot similarly suppose that the welfare vector in the residual state is known. If that were so, r would label an ordinary state. Analysis would remain unchanged when r is added to the universe U. The essence of the difficulty in decision making with the residual state is that $[w(c, r), c \in C]$ is not known.

We observed in the Introduction that research in the economic theory of incomplete contracts attempts to cope by placing a subjective probability distribution on welfare in the residual state. Here we impose the weaker assumption that $[w(c, r), c \in C]$ lies in a specified non-singleton subset of $R^{|C|}$, which we denote $\Omega$.

From the perspective of Bayesian analysis, Ω may be considered the support of a subjective distribution whose shape is not specified. When Ω is a proper subset of $R^{|C|}$, the DM faces a problem with "partially known unknowns." [9]

5.1. Maximization of the Cross-State Sum of Membership-Weighted Welfare

Extending criterion (10) to settings with μ(r) > 0, it is immediate that action c should be weakly preferred to d if

$$(23) \qquad \sum_{s \in U} \mu(s)w(c, s) + \mu(r)\omega(c, r) \geq \sum_{s \in U} \mu(s)w(d, s) + \mu(r)\omega(d, r), \quad \text{all } \omega \in \Omega,$$

or equivalently, if

$$(23') \qquad \sum_{s \in U} \mu(s)[w(c, s) - w(d, s)] \geq \mu(r)[\omega(d, r) - \omega(c, r)], \quad \text{all } \omega \in \Omega.$$

Moreover, c should be strictly preferred to d if the inequality is strict for some ω ∈ Ω.

A definite preference ordering of actions c and d does not exist if (i) ω(d, r) − ω(c, r) varies sufficiently with ω, and (ii) μ(r) is sufficiently large. When this occurs, a DM who wants to maximize the cross-state sum of membership-weighted welfare cannot do so. The DM must resort to some approach for choice under ambiguity.

[9] To restore the classical practice of expressing all uncertainty as lack of knowledge of the true state of nature, one might view each ω ∈ Ω as indexing a different labeled state. Then the concept of a residual state would not be used. Instead, the set of labelled states would be extended from U to (U, Ω). We do not proceed this way because it requires associating a distinct degree of membership with each element of Ω. We find this difficult to contemplate if, as Machina (2003) wrote, the residual state is "a final catch-all state, with a label like 'none of the above,' and a very ill-defined consequence." We admit that a DM may also find it difficult to associate a degree of membership μ(r) with the single residual state that we contemplate.

### 5.2. Membership-Weighted Minimax Regret

It is straightforward to bring the residual state into MMR decision making. Simply extend (11′) to

$$\min_{c \in C} \max \left\{ \max_{s \in U} \mu(s) [\max_{d \in C} w(d, s) - w(c, s)],\ \mu(r) \max_{\omega \in \Omega} [\max_{d \in C} \omega(d, r) - \omega(c, r)] \right\}. \quad (24)$$

Problem (24) augments the space over which regret is maximized to include r. When considering r, one maximizes regret over the space $\Omega$ of feasible welfare functions.

Ceteris paribus, the outcome of (24) depends on $\mu(r)$ and $\Omega$. If $\mu(r)$ is small or the welfare vectors in $\Omega$ imply small regret, maximum regret occurs in the maximization over U as in Section 4. Then the existence of a residual state does not affect decision making. It occurs in the maximization over $\Omega$ if the product of $\mu(r)$ and maximum regret over $\Omega$ is large. Then the behavior of regret across actions in the residual state determines the MMR action.

### 5.3. Bounded Variation Assumptions

Ceteris paribus, the extent of variation in $\omega(d, r) - \omega(c, r)$ over the space $\Omega$ of feasible welfare functions determines whether and how expanding the analysis to the residual state affects decision making. To illustrate, we consider bounded variation assumptions on this differential in the context of a decision problem with two alternatives, e and f. For ease of exposition, suppose e is weakly preferred to f in the universe of labeled states under the membership-weighted extensions of both maximization of subjective expected welfare and minimax regret. The bounded variation assumption is as follows:

$$\delta_{(fe)1} \leq \omega(f, r) - \omega(e, r) \leq \delta_{(fe)u}, \quad \text{all } \omega \in \Omega, \quad (25)$$

for some known constants $\delta_{(fe)1}$ and $\delta_{(fe)u}$, where $\delta_{(fe)1} < \delta_{(fe)u}$

5.3.1. Maximization of the Cross-State Sum of Membership-Weighted Welfare

With e being the weakly preferred action in the universe of labeled states, (10) implies that $\sum_{s \in U} \mu(s)[w(e, s) - w(f, s)] \geq 0$. (23′) implies that e remains weakly preferred in the presence of the residual state if $\sum_{s \in U} \mu(s)[w(e, s) - w(f, s)] \geq \mu(r)\delta_{(fe)u}$ and f is weakly preferred if $\sum_{s \in U} \mu(s)[w(e, s) - w(f, s)] \leq \mu(r)\delta_{(fe)l}$. A definite preference ordering of e and f does not exist if $\mu(r)\delta_{(fe)l} < \sum_{s \in U} \mu(s)[w(e, s) - w(f, s)] < \mu(r)\delta_{(fe)u}$.

5.3.2. Membership-Weighted Minimax Regret

The bounded variation assumption (25) implies that the maximum regret of e in the presence of the residual state is

$$\max \{ \max_{s \in U} \mu(s)[w(f, s) - w(e, s)], \mu(r)\delta_{(fe)u} \}.$$

Noting that (25) may be rewritten as

$$(25') \qquad -\delta_{(fe)u} \leq \omega(e, r) - \omega(f, r) \leq -\delta_{(fe)l}, \quad \text{all } \omega \in \Omega,$$

the maximum regret of f in the presence of the residual state is

$$\max \{ \max_{s \in U} \mu(s)[w(e, s) - w(f, s)], -\mu(r)\delta_{(fe)l} \}.$$

With e being the weakly preferred action in the universe of labeled states, (11′) implies that

$$\max_{s \in U} \mu(s)[w(f, s) - w(e, s)] \leq \max_{s \in U} \mu(s) [w(e, s) - w(f, s)].$$

Thus, e remains the weakly preferred action in the presence of the residual state unless $\mu(r)\delta_{(fe)u}$ is too large. In particular, f is preferred if

$$\mu(r)\delta_{(fe)u} > \max \{ \max_{s \in U} \mu(s)[w(e, s) - w(f, s)], -\mu(r)\delta_{(fe)l} \}$$

Action e is weakly preferred otherwise.

5.3.3. The Precautionary Principle as a Bounded Variation Assumption

Grant and Quiggin (2013) describe a decision problem that implicitly incorporates a bounded variation assumption on welfare in the residual state. They consider choice between two actions that may give rise to the so-called *precautionary principle* favoring "safe options" relative to options that are "subject to unfavorable surprises." They elaborate as follows (p. 751):

> "The case for the precautionary principle arises when a decision maker, such as a regulator, is faced with a choice between alternatives, one of which leads to consequences for which the relevant elements of the state space are well understood and the other which leads to consequences that depend to a significant extent on 'unknown unknowns.' If most surprises are unpleasant, a risk analysis based only on known risks will underestimate costs of choices of the second kind."

In our framework, suppose action e is weakly preferred in the universe of labeled states but is "subject to unfavorable surprises," whereas action f is the "safe option." This suggests that $\delta_{(fe)l}$ is non-negative and $\delta_{(fe)u}$ is relatively large. Then, with membership-weighted maximization of subjective expected welfare, action f is weakly preferred—i.e., the precautionary principle applies—if $\mu(r)\delta_{(fe)l}$ exceeds the membership-weighted difference in welfare in the universe of labeled states. Similarly, with membership-weighted minimax regret, the precautionary principle applies if the weighted maximum regret of e in the residual state, $\mu(r)\delta_{(fe)u}$, exceeds the weighted maximum regret of f in the universe of labeled states.

Loose qualitative reference to the precautionary principle has become common in public discussions of environmental policy, such as a statement released by the European Commission (2000). It has become particularly prevalent to express concern that future global warming may exceed the forecasts of all of the climate models that have been contemplated in analyses of climate policy.

If this concern is warranted, the MMR studies of climate policy that we discussed in Section 2.2 are too optimistic. It would have been more realistic to augment the state space in these studies by a residual state in which each of the policies compared there yields lower welfare than it does in all of the labeled states that were considered. In this context, safe options may be policies that aggressively reduce GHG emission. Policies subject to unfavorable surprises may be ones that reduce emissions gradually.

## 6. Conclusion

We find degree of membership in the state space to be a conceptually appealing way to generalize classical decision theory. We reiterate that the concept differs substantively from but is consistent with the subjective probability distributions of Bayesian decision theory. Bayesian theory places a subjective distribution on the elements of a classical state space, each of whose members has degree of membership one. Degree of membership expresses the distinct idea that a DM may want to quantitatively express uncertainty regarding the composition of the state space. A mathematical distinction between subjective probabilities and degrees of membership is that the former must sum to one, but the latter need not.

We view the primary open question to be the extent to which actual DMs will find degree of membership a useful concept in practice. To explore this, we intend to perform empirical research. We will seek to elicit degrees of membership from actual DMs and learn how they would use them in realistic settings. We will also attempt to learn how DMs combine theory and empirical evidence to form degrees of membership. Among other objectives, we are curious to learn whether DMs use sample data to update degrees of membership in the Bayesian manner or apply the ex ante framework of frequentist statistical theory, wherein sample data should not affect degree of membership.